\documentclass[amsmath,amssymb,twocolumn,showpacs,prb]{revtex4}
\usepackage[dvips]{graphicx}
\usepackage{bm}

\begin{document}
\title{Work fluctuation theorem for a classical circuit coupled to a quantum conductor}

\author{
Y. Utsumi$^1$,
D. S. Golubev$^2$,
M. Marthaler$^{3}$,
Gerd Sch\"on$^{2,3,4}$, 
and
Kensuke Kobayashi$^{5}$
}

\affiliation{
$^1$ Department of Physics Engineering, Faculty of Engineering, Mie University, Tsu, Mie, 514-8507, Japan
\\
$^2$ Institut f\"ur Nanotechnologie, Karlsruhe Institute of Technology, 76021 Karlsruhe, Germany \\
$^3$ Institut f\"{u}r Theoretische Festk\"{o}rperphysik, Karlsruhe Institute of Technology, 76128 Karlsruhe, Germany\\
$^4$ DFG Center for Functional Nanostructures (CFN), Karlsruhe Institute of Technology, 76128 Karlsruhe, Germany \\
$^5$Institute for Chemical Research, Kyoto University, Uji, Kyoto 611-011, Japan
}


\begin{abstract}
We propose a setup for a quantitative test of the quantum fluctuation theorem. 
It consists of a quantum conductor, driven by an external voltage source, and a classical 
inductor-capacitor circuit.
The work done on the system by the voltage source can be expressed by the classical degrees of freedom
of the $LC$ circuit, which are measurable by conventional techniques.
In this way the circuit acts as a classical detector to perform measurements of the quantum conductor. 
We prove that this definition is consistent with the work fluctuation theorem.   
The system under consideration is effectively described by a Langevin equation with 
non-Gaussian white noise. 
Our analysis extends the proof of the fluctuation theorem to this situation.  
\end{abstract}

\date{\today}
\maketitle

\newcommand{\mat}[1]{\mbox{\boldmath$#1$}}

\section{introduction}

The degrees of freedom of physical systems usually fluctuate, with strength which in thermal equilibrium 
 is related to the transport properties by the fluctuation-dissipation theorem. 
Out of equilibrium the fluctuation theorem (FT), or the fluctuation relation, imposes universal constraints on the probability distributions of the fluctuating parameters~\cite{Tobiska,Foerster,SU,Andrieux,Esposito,Campisi}. 
The FT has been studied in a variety of 
contexts, including applications to
electron transport in mesoscopic systems~\cite{GU,Cuetara,Krause,Ganeshan,Averin}. 
There exist several equivalent versions of the FT, all derived from two main assumptions, namely
(i) equilibrium Gibbs form of the initial statistical distribution, and (ii) time reversibility of the microscopic evolution equations.
Below we will focus on one of these versions --- the work FT~\cite{Crooks,Monnai}. 
It is formulated in terms of the distribution $P(W,B)$ of the
work $W$ done on the system by the external force during time $\tau$ in the presence of a magnetic
field $B$. In its simplest form it reads
\begin{eqnarray}
\frac{P(W;B)}{P(-W;-B)}
=
{\rm e}^{\beta W}
\, , 
\label{crooks}
\end{eqnarray}
where $\beta = 1/k_{\rm B} T$ is the inverse temperature. 
It is valid both for classical and for quantum systems~\cite{Campisi}, 
where the magnetic field helps revealing interference effects.
     
The interpretation of the identity (\ref{crooks}) for a classical system is straightforward, 
and there exist no fundamental obstacles to its experimental verification. 
Let us briefly discuss the corresponding experimental procedure. 
One basically needs to switch on an external force at time zero and
switch it off at time $\tau$. During this time interval one 
continuously monitors the change of the relevant system parameters. The work $W$ is usually related
to these parameters in a simple way and  can be computed. Repeating this experiment many times  
one can determine the distribution $P(W;B)$ and verify the identity (\ref{crooks}). In this way 
the fluctuation theorem has been confirmed in various systems ranging from colloidal 
particles in a solution~\cite{Wang} and RNA molecules~\cite{RNA}, 
to quantum dots in the regime of strong Coulomb blockade where individual tunneling 
electrons can be counted~\cite{UG,Kueng}.   

The experimental protocol is more subtle when the object under consideration is a quantum system.
In this case the work $W$ is defined as a difference between the final and initial
energies of the quantum system~\cite{Kurchan,HTasaki,Campisi}. Thus, in order to recover the distribution
$P(W,B)$ one should perform two projective measurements at the
beginning and end of every experimental run. 
While this procedure might work, e.g., for qbits and ultracold atoms\cite{Campisi}, it becomes
 difficult to realize if one deals with more conventional quantum mesoscopic objects like Aharonov-Bohm interferometers or quantum dots.
This is one of the reasons why the FT (\ref{crooks}) has not yet been fully  tested in such systems. 
So far only the relations between the non-linear transport coefficients, 
which follow from the identity (\ref{crooks}) for low bias voltages, have been verified~\cite{Nakamura}.

In order to overcome this problem we propose a different scheme to measure the work $W$. 
On the one hand, it should be applicable in systems involving small mesoscopic conductors. 
On the other hand, as we will show, it is still consistent with the FT (\ref{crooks}).          
Our approach is motivated by the theory of Nazarov and Kindermann~\cite{Nazarov}, who have proposed 
to measure the full counting statistics~\cite{Levitov,NazarovB} (FCS)
of the charge transferred through a quantum conductor with the aid of a classical system coupled to it.
Extending these ideas, we propose to couple the conductor to a classical oscillator made of an inductor $L$ and a capacitor $C$.
To ensure its classical behavior we require the oscillation frequency to be small,
\begin{eqnarray}
\hbar/\sqrt{LC}\ll \max\{T,eV\},
\label{condition}          
\end{eqnarray}
where $V$ is the voltage drop across the quantum conductor.
Since the $LC$ oscillator is a classical system, one can in principle continuously measure the fluctuating voltage $V(t)$ by a sensitive amplifier, 
from which the work $W$ is obtained by classical arguments (see Eq.~(\ref{workexp})). 
This definition of the work would be exact if both the $LC$ circuit and the conductor were classical.
However, it differs from the standard definition of the work in the quantum regime~\cite{Kurchan,HTasaki,Campisi}, and hence
the standard proof of the quantum FT (\ref{crooks}) does not apply any more. We will show that,
 nevertheless, the FT (\ref{crooks}) remains valid under the condition (\ref{condition}). 

Finally, we note that in our model the dynamics of the classical $LC$ oscillator is described by a Langevin equation with white
non-Gaussian noise generated by the quantum conductor. To the best of our knowledge the FT has not yet been proven for this system. 
Thus our analysis is also interesting in this context.

The outline of the paper is as follows: 
In Sec.~\ref{sec:Langevin} we define the model; 
in Sec.~\ref{sec:fcswork} we derive the probability distribution of the work,
show how it is related to the FCS, and how one can treat the back-action of the $LC$-circuit on the conductor; 
in Sec.~\ref{sec:constbias} we show that under constant bias voltage 
the FCS of the work $W$ is equivalent to the FCS of the charge transferred through the quantum conductor; 
in Sec.~\ref{sec:workft} we prove the work FT (\ref{crooks}) for coupled quantum and classical system; 
and in Sec.~\ref{sec:qdab} we apply our theory to a quantum-dot Aharonov-Bohm interferometer. 
Finally, we will summarize our results.

\section{Model}
\label{sec:Langevin}

\begin{figure}
\begin{center}
\includegraphics[width=0.95 \columnwidth]{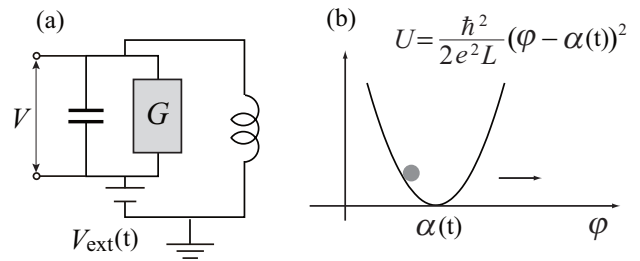}
\caption{
(a) 
Schematics of the system, which consists of a coherent quantum conductor [denoted by $G$] connected to an inductor and a capacitor. 
The voltage drop across the quantum conductor $V$ is measured by a voltmeter. 
(b) A Brownian particle in a driven harmonic potential. 
}
\label{fig:setup}
\end{center}
\end{figure}

We consider the system depicted in Fig.~\ref{fig:setup}(a).  
It consists of a quantum conductor with conductance $G$
coupled in parallel to a capacitor $C$ and in series to an
inductor $L$. A bias voltage $V_{\rm ext}$ is applied 
from the external voltage source. The system is described by the Hamiltonian 
\begin{eqnarray}
\hat H=\hat H_G(\hat p_j,\hat x_j;\hat \varphi)+\hat H_{LC}(\hat q,\hat \varphi;\alpha),
\label{Hfull}
\end{eqnarray}
where $\hat H_G(\hat p_j,\hat x_j;\hat \varphi)$ refers to the conductor
and $\hat H_{LC}(\hat q,\hat \varphi;\alpha)$ to the $LC$ circuit. 
Here $\hat p_j,\hat x_j$ are the degrees of freedom which describe the quantum conductor (they are, for example, the momenta and coordinates of the tunneling electrons, of the electro-magnetic environment, etc.). Our analysis is applicable to a wide range of quantum conductors, and we do not further specify $\hat H_G$. 

The Hamiltonian of the $LC$ circuit reads 
\begin{eqnarray}
\hat H_{LC}(\hat q,\hat \varphi;\alpha)=
\frac{\hat q^2}{2C}+\left( \frac{\hbar}{e} \right)^2\frac{(\hat \varphi-\alpha)^2}{2L}. 
\label{hamiltonian}
\end{eqnarray}
where $\hat q$ is the operator of the charge stored in the capacitor,
related to the voltage drop across the conductor $\hat V$ in a usual way $\hat q=C\hat V$, 
while $\hat \varphi(t)= \int^t dt' e \hat V(t') /\hbar$ 
is the operator of the phase~\cite{AES} associated with the voltage drop $\hat V$, 
while $\alpha(t)= \int^t dt' e V_{\rm ext}(t') /\hbar$ 
is the phase, which characterizes the external voltage bias.
As  discussed, we assume the $LC$-oscillator to be a classical system. 
Hence, one can replace the operators $\hat q,\hat\varphi$
by the classical charge and phase, $q$ and $\varphi$.  

We emphasize that the Hamiltonian (\ref{Hfull}) takes into account the back action of the detector, i.e.\ the $LC$-circuit, on the quantum conductor.
This back action manifests itself through the dependence of the Hamiltonian $\hat H_G$ on the coordinate of the $LC$ oscillator $\hat\varphi$.
Finally, we would like to note that our setup is the electric analog of a colloidal particle dragged by a harmonic optical trap with a velocity $\dot \alpha(t)$~\cite{Wang} [see Fig.~\ref{fig:setup} (b)]. 
However, our environment possesses two striking differences as compared with the colloidal particle case. 
First, the noise is non-Gaussian. 
Second, we can break the time-reversal symmetry by applying a magnetic field~\cite{Sanchez}. 
In general, the probability distribution of noise depends on the direction of this field.

Next, we define the work done by the external voltage source 
on the whole system, i.e., the conductor and the $LC$-circuit, for a given realization of the 
fluctuating time-dependent voltage $V(t)$~\cite{Jarzynski} 
\begin{align}
W[\varphi;\alpha]
&=
\int_{-\tau/2}^{\tau/2} \!\!\! dt \, 
\dot{\alpha}
\frac{\partial H_{LC}(q,\varphi;\alpha)}{\partial \alpha}
\label{work}
\\
&=
\int_{-\tau/2}^{\tau/2} \!\!\! dt \, 
V_{\rm ext}(t)
\int_{-\tau/2}^{t} \!\!\! dt' \, 
\frac{V_{\rm ext}(t')-V(t')}{L}
\, . 
\label{workexp}
\end{align}
Since the $LC$ circuit is classical, the fluctuating voltage $V(t)$, in principle,  can be measured. Hence the work $W[\varphi;\alpha]$ can be 
measured as well. 
Experimentally, one should first record the fluctuating voltage $V(t)$ during a time interval $\tau$. 
Afterwards, the work (\ref{workexp}) can be computed. 
Repeating this measurements many times one can find the probability distribution of the work $P(W,B)$
and then verify the identity (\ref{crooks}). 
This kind of measurements may be challenging at present, but with the development of 
low-invasive and wide-band on-chip electrometers, quantum point contacts 
or single-electron transistors~\cite{Lu,Fujisawa}, such measurements should become possible. 

The problem of measuring the probability distribution of the work $W$ is equivalent
to that of measuring the distribution of the charge transferred through the
conductor, i.e. to the problem of measuring its FCS \cite{Nazarov}. 
The key point here is that the work done on the $LC$-circuit 
turns into Joule heat, which is dissipated in the quantum conductor. 
It suggests that the fluctuation properties of the work, the Joule heat and the transmitted charge are the same. 
One can demonstrate this property from the equations of motion of the circuit 
\begin{eqnarray}
\frac{\hbar}{e}
\dot{\varphi}
&=&
\frac{\partial H_{LC}(\varphi,q;\alpha)}{\partial q}
\, ,
\label{eqm1}
\\
\dot{q}
&=&
-
\frac{e}{\hbar}
\frac{\partial H_{LC}(\varphi,q;\alpha)}{\partial \varphi}
-I(t)
\, ,
\label{eqm2}
\end{eqnarray}
where $I(t)$ is the fluctuating current flowing through the quantum conductor.
The work (\ref{workexp}) is related to the charge 
$Q=\int^{\tau/2}_{-\tau/2} dt' I(t')$ transmitted through the conductor via 
$W = V_{\rm ext}Q - V_{\rm ext}[q(\tau/2)-q(-\tau/2)]$.
In the long-time limit, $\tau\to\infty$, which is relevant under the condition (\ref{condition}), the second term in this expression becomes
much smaller than the first one, which proves our statement.

Eqs. (\ref{eqm1},\ref{eqm2}) are equivalent to the Langevin equation
\begin{eqnarray}
C\frac{\hbar\ddot\varphi}{e} + \frac{\hbar\varphi}{eL} = \frac{V_{\rm ext}t}{L}-I(t).
\label{Langevin}
\end{eqnarray}
In this equation
the fluctuating current $I(t)$ 
plays the role of the noise with a non-zero average value $\bar I$. 
The theory predicts that the correlators of its fluctuations 
$\delta I(t)=I(t) -\bar I$ quickly  decay in time.
For example, the correlator
$ \left\langle \delta I(t_1) \delta I(t_2) \right\rangle  $
decays to zero if $|t_1-t_2|\gg \min\{\hbar/eV,\hbar / T\}$. 
Since the $LC$-oscillator is slow, see  Eq.~(\ref{condition}),
we may consider the currents taken at different times as uncorrelated 
and treat $I(t)$ as white noise. 

In contrast to conventional models, the fluctuations of the current $I(t)$
are not Gaussian. 
We characterize their statistical properties by the probability 
$p(t,Q; B)$ 
that the charge $Q=\int_0^t dt'\,I(t')$ is transferred through the conductor during
time $t$ in the presence of a magnetic field $B$.
It is convenient to introduce the characteristic function (CF) of 
current fluctuations 
\begin{eqnarray}
{\mathcal Z}_{G}(\lambda,V;B)
=
\sum_{Q/e}
{\rm e}^{i \lambda Q/e}
p (t,Q,V;B). 
\label{CFq}
\end{eqnarray}
In the white-noise approximation  considered here
the time dependence of the CF reduces to a simple exponent
\begin{eqnarray}
{\mathcal Z}_G(\lambda,V;B) \approx {\rm e}^{ t {\mathcal F}_G(\lambda,V;B)} ,
\label{Z_long}
\end{eqnarray}
where ${\mathcal F}_G(\lambda,V;B)$ is the cumulant generating function (CGF) of the conductor,
which satisfies the FT ~\cite{Tobiska,SU,Andrieux}, 
\begin{eqnarray}
{\mathcal F}_G(\lambda, V;B)
=
{\mathcal F}_G
(-\lambda+i \beta e V, V;-B). 
\label{ftqc}
\end{eqnarray}

\section{Probability distribution of the work}
\label{sec:fcswork}

We define the CF of the work distribution
\begin{eqnarray}
{\mathcal Z}(\xi; B)
=
\int d W
{\rm e}^{i \xi W}
P(W; B), 
\label{fouriertran}
\end{eqnarray}
and the corresponding CGF
\begin{eqnarray}
{\mathcal F}(\xi)
=
\lim_{\tau \to \infty}
\frac{1}{\tau}
\ln 
{\mathcal Z}(\xi)
\, . 
\label{CGF}
\end{eqnarray}
In order to evaluate the CF (\ref{fouriertran}) we follow the method proposed in  Refs.~\onlinecite{Nazarov,Belzig}. 
We split the measurement time interval $[-\tau/2,\tau/2]$ into $N=\tau/\Delta t$ pieces. 
The time step $\Delta t$ should lie in the range $1/\max\{eV,T\}\lesssim \Delta t \lesssim \sqrt{LC}$, 
i.e., $\Delta t$ is sufficiently short to accurately describe the dynamics of the $LC$ circuit and sufficiently long to allow using the long time approximation for the CF of the quantum conductor (\ref{Z_long}). 
In this case at each time $t_{i}$ the $LC$ circuit and the quantum conductor 
are not entangled, 
and the system density matrix factorizes,
\begin{eqnarray}
\rho(t_i) \approx \rho_{LC}(t_i) \otimes \rho_G(V(t_i)).
\end{eqnarray}  
Here $t_i= i\Delta$ are the discretized times,
$\rho_{LC}(t_i)$ and $\rho_G(V(t_i))$ are
the reduced density matrices of the oscillator and of the quantum conductor, respectively, 
and $V(t_i)$ is the value of the bias voltage during the interval
$t_i<t<t_{i+1}$. 
This voltage drop is induced as the back action of the classical $LC$ circuit.

Next, following Ref.~\onlinecite{Nazarov}, we express the reduced density matrix at time $t_{i}$ in the form 
\begin{eqnarray}
\rho_{LC}(\varphi_{i}^+ , \varphi_{i}^-)
\!\! &=& \!\!
{\rm Tr}
[
\langle \varphi_{i}^+  | \rho(t_{i}) | \varphi_{i}^- \rangle
]
\nonumber \\
\!\! &\approx& \!\!
\int 
{d \varphi_{i-1}^+ }
{d \varphi_{i-1}^-}
\pi_{\Delta t}(
\varphi_{i}^+ ,
\varphi_{i}^-
| 
\varphi_{i-1}^+ ,
\varphi_{i-1}^-
;
\alpha_i
)
\nonumber \\ && \times 
\rho_{LC}(\varphi_{i-1}^+ , \varphi_{i-1}^-), 
\label{dm}
\end{eqnarray}
where the propagator for one time step $\pi_{\Delta t}$ reads
\begin{align}
& \pi_{\Delta t}(
\varphi_{i}^+ ,
\varphi_{i}^-
| 
\varphi_{i-1}^+ ,
\varphi_{i-1}^-
;
\alpha_i
)
=
\int 
\frac{d q_{i}^+ }{2 \pi e} 
\frac{d q_{i}^-}{2 \pi e} 
\nonumber \\ &\times 
{\rm e}^{
i q_{i}^+  (\varphi_{i}^+ -\varphi_{i-1}^+ )/e
-
i q_{i}^- (\varphi_{i}^- -\varphi_{i-1}^-)/e
}
\nonumber \\ & \times 
{\rm e}^{
-i 
[
H_{LC}(\varphi_{i}^+ , q_{i}^+ ; {\alpha}_{i})
-
H_{LC}(\varphi_{i}^-, q_{i}^-; {\alpha}_{i})
]
\Delta t /\hbar
}
\nonumber \\ & \times 
{\rm Tr}
\bigg[
{\rm e}^{-iH_G(\varphi_{i}^{+};B) \Delta t/\hbar}
\rho_G(V(t_{i-1}))
{\rm e}^{iH_G(\varphi_{i}^{-};B) \Delta t/\hbar}
\bigg], 
\label{pro}
\end{align}
with
$\alpha_i=\alpha(t_i)$, 
etc. 
The operator of the current through the conductor is related to its
Hamiltonian as follows: 
$
\hat{I}
=
(e/i \hbar)
\partial H_G(\varphi;B)/
\partial \varphi
|_{\varphi=0}
$. 

In the Keldysh formalism $\varphi=(\varphi^+ +\varphi^-)/2$
and $q=(q^+ +q^-)/2$
are related to classical dynamical variables, which are measurable, 
while 
$\tilde{\varphi}=\varphi^+ -\varphi^-$
and
$\tilde{q}=q^+ -q^-$
are `quantum' variables, which are small in the classical limit.
We perform a first-order expansion in $\tilde{\varphi},\tilde{q}$, approximating
the difference of the Hamiltonians as  
$
H_{LC}(\varphi^{+}, q^{+}; {\alpha})
-
H_{LC}(\varphi^-, q^-; {\alpha})
\approx
(
\tilde{\varphi} 
\partial_\varphi
+
\tilde{q} 
\partial_q
)
H_{LC}(\varphi, q; {\alpha})
$. 
Furthermore, we define the free energy of the classical $LC$ circuit 
\begin{eqnarray}
F_{LC}(\alpha)
=
- k_{\rm B} T
\ln
\int \frac{d \varphi d q}{2 \pi e}
\exp \left[-\beta H_{LC}(\varphi,q;\alpha) \right]. 
\end{eqnarray}
Finally, the CF~(\ref{fouriertran}) may be transformed to the form 
\begin{widetext}
\begin{align}
{\mathcal Z}
=
\left \langle {\rm e}^{i \xi \, {W}[\varphi;\alpha] } \right \rangle
 \equiv &
\lim_{N \to \infty}
\int 
\frac{d \varphi_0 d q_0}{2 \pi e}
{\rm e}^{-\beta [H_{LC}(\varphi_0,q_0;\alpha_0)-F_{LC}(\alpha_0)]}
\prod_{i=1}^N
\int d \varphi_i d \tilde{\varphi}_i
{\rm e}^{i \xi 
(\alpha_i-\alpha_{i-1}) \, 
{\partial H_{LC}({\varphi}_i,{q}_i;{\alpha}_i)}/{\partial \alpha}
}
\nonumber \\ & \times
\pi_{\Delta t}(
\varphi_{i}+\tilde{\varphi}_{i}/2,
\varphi_{i}-\tilde{\varphi}_{i}/2
| 
\varphi_{i-1}+\tilde{\varphi}_{i-1}/2,
\varphi_{i-1}-\tilde{\varphi}_{i-1}/2
;
\alpha_i
)
\label{defz}
\\ = &
\lim_{N \to \infty}
\int 
\frac{d \varphi_0 d q_0}{2 \pi e}
{\rm e}^{-\beta [H_{LC}(\varphi_0,q_0;\alpha_0)-F_{LC}(\alpha_0)]}
\left(
\prod_{i=1}^N
\int \frac{d q_i d \varphi_i}{2 \pi e}
\int \frac{d \tilde{q}_i d \tilde{\varphi}_i}{2 \pi e}
\right)
{\rm e}^{i S_{\rm t}/\hbar}
. 
\label{fcsw2}
\end{align}
\end{widetext}
Eq.~(\ref{defz}) is interpreted as follows. As in a real experiment, 
the  classical phase $\varphi_i$ is supposed to be measured at every time $t_i$.    Then   
the derivative ${\partial H_{LC}}/{\partial \alpha}$, which is independent of the charge $q$, may be computed.
Next the exponent 
${\rm exp}[i \xi (\alpha_i-\alpha_{i-1}) {\partial H_{LC}({\varphi}_i,{q}_i;{\alpha}_i)}/{\partial \alpha}]$ 
is constructed and averaged over all possible realizations of the current fluctuations. The latter are
described by the propagators $\pi_{\Delta t}$ coming from the evolution of the quantum conductor.

In Eqs. (\ref{fcsw2}) we have introduced the action of the whole system $S_{\rm t}$, which is composed of three parts,
\begin{eqnarray}
S_{\rm t} 
&=&
\xi \hbar W
+S_{LC}
+
S_G. 
\label{action1}
\end{eqnarray}
Here, $W$ is the discretized version of the work (\ref{work}), 
\begin{eqnarray}
W[\{ \varphi_i,\alpha_i \} ]=
\sum_{i=1}^{N}
(\alpha_i-\alpha_{i-1}) 
\frac{\partial H_{LC}({\varphi}_i,{q}_i;{\alpha}_i)}{\partial \alpha}, 
\label{actionwork}
\end{eqnarray}
and $S_{LC}$ is discrete form of the Martin-Siggia-Rose action~\cite{Kamenev} of the $LC$ circuit 
\begin{eqnarray}
S_{LC}
&=&
\sum_{i=1}^N
\tilde{q}_i
\left(
\frac{\hbar}{e}
\frac{\varphi_i-\varphi_{i-1}}{\Delta t}
- 
\frac{\partial H_{LC}({\varphi}_i,{q}_i;{\alpha}_i)}{\partial q}
\right)
\Delta t
\nonumber \\
&&
+ \, 
\tilde{\varphi}_i
\left(
-
\frac{\hbar}{e}
\frac{q_{i+1}-q_{i}}{\Delta t}
-
\frac{\partial H_{LC}({\varphi}_i,{q}_i;{\alpha}_i)}{\partial \varphi}
\right)
\Delta t 
\nonumber \\
&&
+ \, 
(\hbar /e)
(
q_{N+1}
\tilde{\varphi}_{N}
-
q_{1}
\tilde{\varphi}_{0}
).
\label{actionLC}
\end{eqnarray}
In what follows we will omit unimportant boundary terms in the last of this expression.  
Finally, the action of the conductor takes the form
\begin{eqnarray}
i\frac{S_G}{\hbar} =
\sum_{i=1}^N
\Delta t \, 
{\mathcal F}_G
\left(
-\tilde{\varphi}_{i} ,
\frac{\hbar (\varphi_i-\varphi_{i-1})}{e \Delta t};B
\right)
\, , 
\label{actionqc}
\end{eqnarray}
where
\begin{eqnarray}
{\mathcal F}_G
(\lambda,V;B)
&=&
\lim_{t \to \infty}
\frac{1}{t}
\ln 
{\rm Tr}
\left[ 
{\rm e}^{-i H_G(-\lambda/2;B) t/\hbar}
\right.
\nonumber \\ && \times
\left. 
\rho_q(V) \, 
{\rm e}^{i H_G(\lambda/2;B) t/\hbar}
\right]
\, , 
\label{cgf_q}
\end{eqnarray}
is the standard CGF of a quantum conductor~\cite{Levitov,NazarovB}. 

In order to demonstrate the equivalence of the abstract formulation  of the problem
in terms of the CF (\ref{fcsw2}) to the Langevin equation approach (\ref{eqm1},\ref{eqm2}),
we evaluate the integrals over $\tilde{q}_i$ and $\tilde{\varphi}_i$ in Eq. (\ref{fcsw2}) and 
transform it to the form 
\begin{eqnarray}
{\mathcal Z}
&=&
\lim_{N \to \infty}
\int 
\frac{d \varphi_0 d q_0}{2 \pi e}
{\rm e}^{-\beta [H_{LC}(\varphi_0,q_0;\alpha_0)-F_{LC}(\alpha_0)]}
\nonumber \\ && \times
\left(
\prod_{i=1}^N
\int {d q_i d \varphi_i}
\sum_{\Delta Q_i/e}
\right)
{\rm e}^{i \xi W}
\nonumber \\ && \times
\bigg[
\prod_{i=1}^N
\delta \!
\left(
{\varphi_i-\varphi_{i-1}}
-
\frac{e}{\hbar}
\frac{\partial H_{LC}({\varphi}_i,{q}_i;{\alpha}_i)}{\partial q}
\Delta t
\right)
\nonumber \\
&&
\times
\delta \!
\left(
-
q_{i+1}+q_{i}
-
\Delta Q_i
-
\frac{e}{\hbar}
\frac{\partial H_{LC}({\varphi}_i,{q}_i;{\alpha}_i)}{\partial \varphi}
\Delta t
\right)
\nonumber \\ && \times
p(\Delta t,\Delta Q_i,\hbar (\varphi_i-\varphi_{i-1})/(e \Delta t);B)\bigg].
\label{fcsw1} 
\end{eqnarray}
This expression is nothing else but the representation of the discrete Langevin equation in the presence of the non-Gaussian white noise $\Delta Q$.
It is easy to see that these equations become equivalent to Eqs.~(\ref{eqm1},\ref{eqm2}) in the 
limit $N\to\infty$, $\Delta t\to 0$.

Eq.\ (\ref{fcsw2}) is the main result of this section and provides an exact formal expression for the CF 
of a system governed by Langevin equations with non-Gaussian white noise, see Eqs.\ (\ref{eqm1},\ref{eqm2}) and (\ref{Langevin}). 
The quasi-stationary approximation, which we have used above,
has been used earlier to analyze the properties of the Josephson junction threshold detectors~\cite{Tobiska1}.  
It is also very similar to the stochastic path-integral approach~\cite{Pilgram}.

\section{Saddle-point approximation under constant bias voltage}
\label{sec:constbias}

Let us consider the effect of a constant bias voltage, $V_{\rm ext}=$const.  
In the limit of sufficiently long measurement time $\tau$ 
we may use the saddle-point approximation to evaluate the integral (\ref{fcsw2}). 
Considering the limit  $N \! \to \! \infty$, 
we solve the equations 
$\delta S_t/\delta \varphi(t)=
\delta S_t/\delta \tilde{\varphi}(t)=
\delta S_t/\delta q(t)=
\delta S_t/\delta \tilde{q}(t)=0$.
The corresponding solution reads: 
$\tilde{q}(t)
=
\dot{q}(t)=
0$, 
$\varphi(t)=eV_{\rm ext}t/\hbar+\varphi(0)$ 
and 
$\tilde{\varphi}(t)=- \xi eV_{\rm ext}$. 
In this approximation the CGF of the work (\ref{CGF}) acquires a simple form  
\begin{eqnarray}
{\mathcal F}(\xi)
& \approx &
\frac{i}{\hbar}
(\tilde{\varphi}+ \xi eV_{\rm ext})
\frac{\partial H_{LC}}{\partial \varphi}
+
{\mathcal F}_G(\xi e V_{\rm ext},V_{\rm ext};B)
\nonumber \\
&=&
{\mathcal F}_G(\xi e V_{\rm ext},V_{\rm ext};B)
\, . 
\label{fcswq}
\end{eqnarray}
It is interesting that in this regime the contributions $S_{LC}$ and $\hbar \, \xi W$ in the total action (\ref{action1}) cancel each other.
Thus we have proven that the statistical properties of the work done on the classical $LC$-circuit 
and those of the current flowing through the quantum conductor are the same. 
This interesting conclusion remains valid only in the saddle-point approximation, 
which works well as long as $1/(L C^2 R) \ll 1$, 
where $R$ is the resistance of the quantum conductor. 

By virtue of the FT (\ref{ftqc}), which is valid for an isolated conductor, the CGF of the work satisfies
\begin{eqnarray}
{\cal F}(\xi; B)={\cal F}(-\xi+i \beta; -B).
\label{FT_CGF}
\end{eqnarray}
This identity is equivalent to the work FT~(\ref{crooks}).

\section{Work fluctuation theorem for coupled classical and quantum systems}
\label{sec:workft}

In this section we show that the FT (\ref{FT_CGF}) holds even beyond the saddle-point
approximation as long as one uses the quasi-stationary approximation introduced in Sec. (\ref{sec:fcswork}).  
The basis of our proof is the FT (\ref{ftqc}) for the charge transport through the quantum conductor. 
As a first step we apply the FT (\ref{ftqc}) $N$ times for every time interval $t_i<t<t_{i+1}$.
Since the quantum phase $- \tilde\varphi_i$ and the combination 
$\hbar(\varphi_i-\varphi_{i-1})/e\Delta t$ 
play the same role in the action (\ref{actionqc}) as the counting field $\lambda$ and the bias voltage $V$ in the Eq. (\ref{ftqc}), respectively, the transformation $
\lambda\to -\lambda+i\beta V$ 
in Eq. (\ref{ftqc}) translates into the replacement 
$-\tilde\varphi_i\to \tilde\varphi_i+i\beta\hbar(\varphi_i-\varphi_{i-1})/e\Delta t$. 
Similarly, we should replace the quantum charge $-\tilde q_i$ with $\tilde q_i+i\beta\hbar(q_i-q_{i-1})/e\Delta t$. 
At the next step we invert the sings of the quantum phase and charge. 
Combining these two operations we arrive at the following transformation in Eq. (\ref{fcsw2})
\begin{eqnarray}
\tilde{\varphi}_i
&\to&
\tilde{\varphi}_i
-
i \beta \hbar
(\varphi_i-\varphi_{i-1})/\Delta t
\nonumber
\\
\tilde{q}_i
&\to&
\tilde{q}_i
-
i \beta \hbar
(q_i-q_{i-1})/\Delta t,
\label{transformation} 
\end{eqnarray}
($i=1, \cdots, N$). 
One can show that its Jacobian equals to $1$. 
Under the transformation (\ref{transformation}) the action of the quantum conductor (\ref{actionqc}) acquires the form 
\begin{align}
{S_G}
\to
- i \hbar
\sum_{i=1}^N
\Delta t \, 
{\mathcal F}_G
\left(
\tilde{\varphi}_i
, 
\frac{(\varphi_i-\varphi_{i-1}) \hbar}{e \Delta t}
;-B
\right) .
\label{trsq}
\end{align}
Likewise, the action for the $LC$-circuit (\ref{actionLC}) becomes
\begin{eqnarray}
S_{LC}
& \to &
S_{LC}
+
i \beta \hbar \, Q_h 
\, ,
\\
Q_h &=&
\sum_{i=1}^N
\left[
(q_i-q_{i-1})
\frac{\partial H_{LC}({\varphi}_i,{q}_i;{\alpha}_i)}{\partial q}
\right. 
\nonumber \\ && 
\left. 
+
(\varphi_i-\varphi_{i-1})
\frac{\partial H_{LC}({\varphi}_i,{q}_i;{\alpha}_i)}{\partial \varphi}
\right]
\, , 
\end{eqnarray}
where we neglected irrelevant terms. 
The combination $Q_h$ may be interpreted as the heat absorbed by the quantum conductor. 
With its aid the first law of thermodynamics, or energy conservation,
may be written in the form  
$$
H_{LC}(\varphi_N,q_N;\alpha_N)
-
H_{LC}(\varphi_0,q_0;\alpha_0)
\approx
Q_h + {W} \, ,
$$
and thus, we find
\begin{eqnarray}
&& S_{LC}
\to
S_{LC}
+
i \beta \hbar
\, \left[
H_{LC}(\varphi_N,q_N;\alpha_N)
\right. 
\nonumber \\
&&
\left.
-H_{LC}(\varphi_0,q_0;\alpha_0)
-{W}
\right]
\, .  
\label{trslc}
\end{eqnarray}

Next we perform the time-reversal operation 
$t \to -t$, 
$q \to -q$ and 
$\tilde{q} \to -\tilde{q}$. 
Under this transformation the external driving is reversed 
and the phase $\alpha(t)$ is replaced by a time reversed one, $\alpha_R(t)=\alpha(-t)$. 
In the discrete form this transformation reads 
\begin{eqnarray}
&&
\tilde{\varphi}_{i} \to \tilde{\varphi}_{N-i} \, ,
\;\;\;\;
\varphi_i \to \varphi_{N-i} \, ,
\\ &&
\tilde{q}_{i} \to - \tilde{q}_{N-i+1} \, ,
\;\;\;\;
q_i \to -q_{N-i+1} \, ,
\end{eqnarray}
and 
$\alpha_{N-j+1} = \alpha_{R \, j}$. 
Keeping in mind the properties of the Hamiltonian, 
$
H_{LC}(\varphi,q;\alpha)
=
H_{LC}(\varphi,-q;\alpha)
$
and 
$\partial H_{LC}(\varphi,q;\alpha)/\partial q
=-\partial H_{LC}(\varphi,-q;\alpha)/\partial q$, 
we arrive at the following transformations, 
up to ${\mathcal O} (1)$,  
\begin{eqnarray}
&& {W} \to -{W}_R \, ,
\\
&& S_G \to S_{G,R} \, ,
\\
&& S_{LC}
+ i \beta \hbar \, 
H_{LC}(\varphi_0,q_0;\alpha_0)
\nonumber \\
&\to&
S_{LC ,R}
+i \beta \hbar \, H_{LC}(\varphi_0,\varphi_0,\alpha_{R,0})
+i \beta \hbar \, {W}_R
\, , 
\end{eqnarray}
where we assumed 
$\alpha_{i}-\alpha_{i-1} \propto \Delta t$. 
${W}_R$ and $S_{LC,R}$ are obtained from 
${W}$ (\ref{actionwork}) and $S_{LC}$ (\ref{actionLC}) by means of the replacement 
$\alpha \to \alpha_R$. 
The action of the conductor is transformed as follows 
\begin{align}
{ S_{G,R} }
=
- i \hbar
\sum_{j=1}^N
\Delta t \, 
{\mathcal F}_G
\! \left(
\tilde{\varphi}_{j}
, 
-\frac{(\varphi_{j}-\varphi_{j-1}) \hbar}{e \Delta t}
;-B
\right)
. 
\end{align}
Note that the second argument of the CGF, i.e. the voltage drop, changes its sign. 
It indicates, in turn, that the source and drain electrodes of the quantum conductor 
are effectively interchanged after the time reversal. 

After all these manipulations, we can derive the following identity 
\begin{eqnarray}
\left \langle {\rm e}^{i \xi \, {W} } \right \rangle
\! &=& \! 
\lim_{N \to \infty}
\int 
\frac{d \varphi_0 d q_0}{2 \pi e}
{\rm e}^{-\beta [H_{LC}(\varphi_0,q_0;\alpha_{R0})-F_{LC}(\alpha_{R0})]}
\nonumber \\ && \times
\left(
\prod_{i=1}^N
\int \frac{d q_i d \varphi_i}{2 \pi e}
\int \frac{d \tilde{q}_i d \tilde{\varphi}_i}{2 \pi e}
\right)
\nonumber \\ && \times
{\rm e}^{i (-\xi+i \beta) \tilde{W}_R
+
i
[
S_{LC , R}
+
S_{G ,R}
]/\hbar
}
\nonumber \\ && \times
{\rm e}^{-\beta [F_{LC}(\alpha_N)-F_{LC}(\alpha_0)]}
\\
&=&
\left \langle {\rm e}^{i (-\xi+i \beta) {W}_R
} \right \rangle_R
{\rm e}^{\beta [F_{LC}(\alpha_0)-F_{LC}(\alpha_{N})]} \, , 
\label{workft1}
\end{eqnarray}
which is written in an equivalent form
\begin{eqnarray}
{\mathcal Z}(\xi)
=
{\rm }^{-\beta [F_{LC}(\alpha(\tau/2))-F_{LC}(\alpha(-\tau/2))]}
{\mathcal Z}_R(-\xi+i \beta). 
\label{workft2}
\end{eqnarray}
After Fourier transformation we arrive at the work FT 
\begin{eqnarray}
\frac{P(W)}{P_R(-W)}
=
{\rm e}^{\beta [F_{LC}(\alpha(-\tau/2))-F_{LC}(\alpha(\tau/2))] + \beta W}, 
\label{crooks_1}
\end{eqnarray}
which is more general than form (\ref{crooks}) quoted in the introduction
and is applicable for time-dependent bias voltages $V_{\rm ext}(t)$.
The subscript $R$ in Eqs. (\ref{workft2},\ref{crooks_1}) indicates the
time reversal operation. The latter consists of three steps: 
(i) interchanging of the source and drain electrodes of the quantum conductor, 
(ii) replacement of $\alpha(t)$ with $\alpha_R(t)=\alpha(-t)$, 
and 
(iii) reversal of the magnetic field $B \to -B$.
This completes the proof of the FT in general case.

The general time reversal operation described above may
be difficult to realize in experiment. Fortunately, 
it may be simplified in many cases. Consider, for example,
the model introduced in Sec.\ II. Since the Hamiltonian
of the $LC$ circuit has the symmetry $H_{LC}(\varphi,q;\alpha)=H_{LC}(-\varphi,-q;-\alpha)$,
one can perform an additional transformation 
$\varphi_i\to-\varphi_i$,
$\tilde{\varphi}_i\to - \tilde{\varphi}_i$, 
$q_i\to -q_i$, 
$\tilde{q}_i\to -\tilde{q}_i$ 
in  Eq. (\ref{workft1}),
which results in the following identity
\begin{eqnarray}
{\mathcal Z}(\tau,\xi,B;\alpha(\tau')) = {\mathcal Z}(\tau,-\xi+i\beta,-B;-\alpha(-\tau')). 
\label{workft3}
\end{eqnarray}
Here we have also used the fact that the free energy of the $LC$ oscillator
does not depend on $\alpha$ and hence $F_{LC}(\alpha(-\tau/2))-F_{LC}(\alpha(\tau/2))\equiv 0$.   
Next, if the external bias voltage is constant, then $\alpha(\tau')=-\alpha(-\tau')=eV_{\rm ext}\tau'$,
and the FT (\ref{workft3}) becomes equivalent to the Eq.\ (\ref{crooks}).
Thus, in order to perform the time reversal in this system experimentally, one just  
needs to change the sign of the magnetic field.  

The simplified version of the FT  
(\ref{crooks}) is also valid if the quantum conductor has an antisymmetric
$I$-$V$ curve, $I(-V)=-I(V)$. 
More precisely, it is valid when the CGF of the conductor satisfies the symmetry 
\begin{eqnarray}
{\cal F}_G(\lambda,V ; B)={\cal F}_G(-\lambda,-V ; B), 
\end{eqnarray}
and the Eq. (\ref{workft1}) reduces to Eq. (\ref{workft3}) regardless
of the symmetries of the Hamiltonian $H_{LC}$.

We conclude this section with two  remarks. 
First, we would like to emphasize once again that our approach takes into account the back action of the $LC$ circuit on the quantum conductor.
Moreover, this back action is essential to ensure the validity of the FT. 
Second, our analysis may also be interpreted as
the proof of the FT for a Langevin equation with
non-Gaussian white noise (\ref{Langevin}), thus  extending the existing 
proof of the FT for the Langevin equation with Gaussian noise~\cite{Kurchan1}.

\section{Quantum-dot Aharonov-Bohm interferometer}
\label{sec:qdab}

\begin{figure}[ht]
\begin{center}
\includegraphics[width=0.50 \columnwidth]{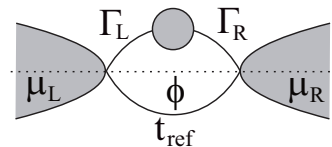}
\caption{
Aharonov-Bohm interferometer embedded with a quantum dot in one arm.
The magnetic flux $\Phi$ threads through the ring and electron wave function acquire the AB phase $\phi$ once it travels in the clockwise direction. 
}
\label{fig:ab}
\end{center}
\end{figure}

In this section we illustrate our results by applying them to an Aharonov-Bohm (AB) interferometer 
with a quantum dot (QD) embedded in one of its arms~\cite{US,Koenig,Hofstetter} [Fig.~\ref{fig:ab}]. 
In this setup the Coulomb interaction and the magnetic field induce asymmetry in the nonequilibrium current distribution~\cite{Sanchez}. 
The microscopic theory of this system based on an extended Anderson model 
has been developed in Ref.~\onlinecite{US}. 
Here, we briefly summarize its key points. 

The $S$-matrix of the QD AB ring~\cite{US,Koenig,Hofstetter}, 
\begin{eqnarray}
\mat{ S }(E ;\phi)
=
\left(
\begin{array}{cc}
S_{LL}(E ;\phi) & S_{LR}(E ;\phi) \\
S_{RL}(E ;\phi) & S_{RR}(E ;\phi)
\end{array}
\right) 
\, ,
\end{eqnarray}
satisfies the micro-reversibility, 
$S_{r r'}(E; \phi)=S_{r' r}(E; -\phi)$. 
Its four components read 
\begin{eqnarray}
S_{LL/RR}
&=&
1
-
\frac{
i \Gamma_{LL/RR}
+
t_{\rm ref}
\sqrt{\Gamma_L \Gamma_R}
\cos \phi
+
t_{\rm ref}^2 \, 
E/2
}
{\Delta(E ;\phi)}
\, ,
\nonumber \\
\\
S_{RL/LR}
&=&
-
i
\frac{
{\rm e}^{\pm i \phi}
t_{\rm ref}
\, 
E
+
\sqrt{\Gamma_L \Gamma_R}
}
{\Delta(E ;\phi)}
\, ,
\\
\Delta
&=&
\frac{
t_{\rm ref}
\sqrt{\Gamma_L \Gamma_R}
\cos \phi
}{2}
+
\left(
1+\frac{t_{\rm ref}^2}{4}
\right)
E
+
i
\frac{\Gamma}{2}
\, , 
\end{eqnarray}
where we set the dot energy level as $\epsilon_{\rm D}=0$. 
Here $\Gamma_{L/R}$ are the tunnel couplings between the quantum dot and the left/right lead, 
$\Gamma=\Gamma_L+\Gamma_R$. 
An electron can also be transmitted through the lower reference arm, characterized by the tunneling amplitude $t_{\rm ref}$. The AB phase,
$
\phi
=2 \pi \, \Phi/\Phi_0
$
is given by the ratio of the magnetic flux $\Phi$ threading the ring and the flux quantum
$\Phi_0=hc/e$. It acquires a minus sign when the magnetic field is reversed 
$\phi(B)=-\phi(-B)$.

\begin{figure}[ht]
\begin{center}
\includegraphics[width=0.9 \columnwidth]{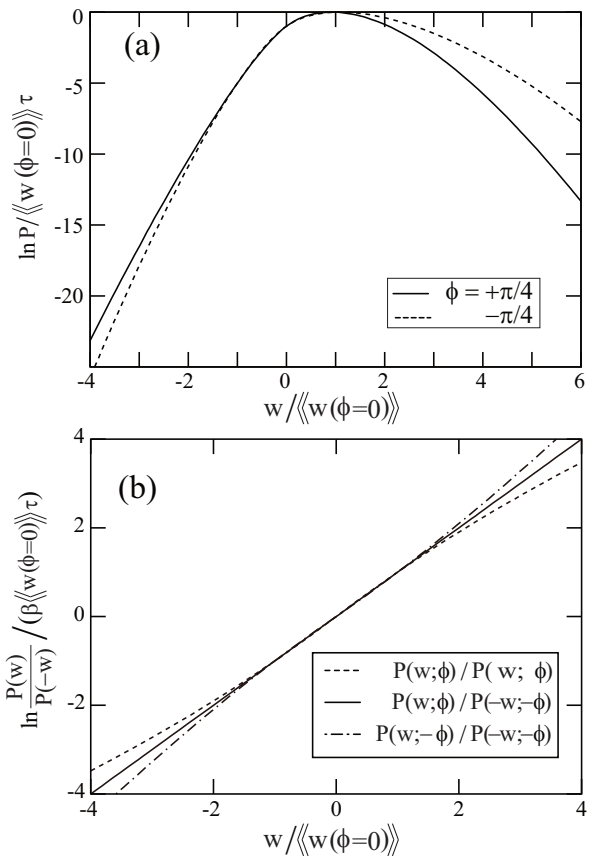}
\caption{
(a) Probability distributions of work for $\phi=\pm \pi/4$. 
(b) Ratio between the positive and negative work probability distributions. 
When the direction of magnetic field is also reversed, the steady-state work fluctuation theorem is satisfied.  
The parameters are
$U/\Gamma=4$, 
$V=\Gamma$, 
$k_{\rm B} T=0.2 \Gamma$, 
$\Gamma_L=0.25 \Gamma$, 
$\Gamma_R=0.75 \Gamma$, 
and 
$
t_{\rm ref}=0.25$, 
and 
$\kappa_L=-\kappa_R=0.5$. 
The average is 
$\langle\!\langle w(\phi=0) \rangle\!\rangle/(M eV_{\rm ext}) 
\approx 2.9 \times 10^{-2} \Gamma/\hbar$. 
}
\label{fig:plot}
\end{center}
\end{figure}

Within the mean-field approximation for the on-site Coulomb interaction $U$, 
the CGF of the AB interferometer is given by the following expression~\cite{US}, 
\begin{eqnarray}
{\mathcal F}_G
(V,\lambda;B)
=
{\mathcal F}_{\rm AB}
(V,\lambda,v_c,v_q;B)
-
M
v_c v_q/U
\, . 
\label{cgfab}
\end{eqnarray}
Here $M$ indicates the degeneracy including channel and spin. 
The mean-field approximation is correct in the limit of $M \to \infty$. 
The CGF for the QD AB ring is, 
\begin{eqnarray}
{\mathcal F}_{\rm AB}
&=&
\frac{M}{2 \pi}
\int d \omega
\ln 
\det \! 
\left[
\mat{1}
+
\mat{f}
\mat{K}
\right]
\, ,
\\
\mat{K}
&=&
{\rm e}^{i \mat{\lambda}/2} \, 
\mat{S}(E - v_c - i v_q/2)^\dagger \, 
{\rm e}^{-i \mat{\lambda}} \, 
\mat{S}(E - v_c - i v_q/2) \, 
\nonumber \\ && \times
{\rm e}^{i \mat{\lambda}/2}
-
\mat{1}
\, , 
\end{eqnarray}
where  
$\mat{1}$
is a $2 \times 2$ unit matrix, 
$
\mat{\lambda}
={\rm diag}(\lambda,0)
$
and 
$
\mat{f}
={\rm diag}(
f(E-\kappa_L eV), 
f(E-\kappa_R eV)
)
$
is the matrix of the Fermi distribution function 
$f(E)
=
1/(\exp(\beta E)+1)$. 
The two parameters, $v_q$ and $v_c$, are determined from the coupled saddle-point equations, 
\begin{eqnarray}
v_q=\frac{U}{M}\frac{\partial {\mathcal F}_{\rm AB} }{\partial v_c}
\, ,
\;\;\;\;
v_c=\frac{U}{M}\frac{\partial {\mathcal F}_{\rm AB} }{\partial v_q}
\, . 
\end{eqnarray}

Returning to the work fluctuation theorem, we note that 
in the limit of long measurement time $\tau$ 
it is more convenient to define the power $w=W/\tau$ instead of the work. 
Applying the method described in the previous section to this system,
and making use of the saddle-point approximation also for the inverse Fourier transform, 
\begin{eqnarray}
P(w)
\approx
\frac{1}{2 \pi}
\int d \xi 
{\rm e}^{-i \tau  w \xi + \tau {\mathcal F}(\xi) }
\, , 
\end{eqnarray}
with ${\mathcal F}$ given by~(\ref{fcswq}), 
we obtain the distribution function $P(w)$ in the form
\begin{eqnarray}
\ln P(w)/\tau
&\approx&
{\mathcal F}_G(\xi^* e V_{\rm ext} ,V_{\rm ext} ;B)-i \, \xi^* w
\, ,
\label{ab1}
\\
w 
&=&
\frac{ \partial }{\partial (i \xi^*)}
{\mathcal F}_G(\xi^* e V_{\rm ext} ,V_{\rm ext} ;B)
\, . 
\label{ab2}
\end{eqnarray}

Figure \ref{fig:plot}(a) shows the probability distributions of the work for negative and positive
values of the AB phase. For the chosen parameters
they are both non-Gaussian and differ significantly when the direction of the magnetic field
is reversed. 
Figure \ref{fig:plot}(b) shows the ratio between the probability distributions for positive
 and negative work. The solid line, obtained with appropriate change of the sign of the magnetic field
 satisfies the work FT. For comparison we also show the ratios when
 the magnetic field is not reversed, in which case the work FT would not be satisfied 
(dashed and dot-dashed lines).

\section{Summary}
\label{sec:summary}

We have proposed an experimental setup which may be used to test the quantum fluctuation theorem. 
It consists of the quantum conductor coupled to a classical $LC$ circuit. 
We note that the usual definition of the work done by an external force on a quantum system~\cite{Kurchan,HTasaki,Campisi} is not convenient when applied to  transport experiments in mesoscopic structures.
Therefore we propose an alternative definition of the work (\ref{workexp}) by expressing it through the degrees of freedom of a classical $LC$ oscillator, which may be measured by conventional techniques.
Our approach takes into account the back action of the $LC$-circuit on the quantum conductor.  
We have proven the work fluctuation theorem for this system and shown 
that under constant bias voltage and with properly chosen parameters of the $LC$ circuit, the probability distribution of the work is directly related to the probability distribution of current flowing through the quantum conductor.  
We applied our theory to the quantum-dot Aharonov-Bohm interferometer and demonstrated the magnetic field induced asymmetry in the work distribution. 
We expect that the probability distribution of the work can be measured with  
currently developed ultra-fast and ultra-sensitive on chip electrometers, such as single-electron transistors or quantum point contacts.
Finally, the classical system coupled with the quantum conductor is effectively described by a Langevin equation with non-Gaussian white noise. 
Therefore our analysis also extends the proof of the fluctuation theorem to this situation.

We thank 
Toshimasa Fujisawa, 
Hisao Hayakawa, 
Bruno K\"ung
and 
Keiji Saito for helpful discussions.
This work has been partially supported by Strategic International Cooperative Program of the Japan Science and Technology Agency (JST) and by the German Science Foundation (DFG), the Okasan-Katoh Foundation, Grant-in-Aid for Young Scientists (B) (No.23740294) and Young Researcher Overseas Visits Program for
Vitalizing Brain Circulation (R2214) from the JSPS.

\end{document}